\documentclass[aps,pra,twocolumn,superscriptaddress,showpacs]{revtex4}

\usepackage{graphicx}

\usepackage{array}

\def\be{\begin{equation}}
\def\ee{\end{equation}}
\def\bea{\begin{eqnarray}}
\def\eea{\end{eqnarray}}
\def\ba{\begin{array}}
\def\ea{\end{array}}
\def\bdm{\begin{displaymath}}
\def\edm{\end{displaymath}}

\begin{document}



\title{SU($N$) Fermi liquid at finite temperature
\\
}

\author{Chi-Ho Cheng}
\affiliation{Department of Physics,
National Changhua University of Education,
Changhua 500, Taiwan}
\author{S.-K. Yip}
\affiliation{Institute of Physics, Academia Sinica, Taipei 115, Taiwan}
\affiliation{Institute of Atomic and Molecular Sciences, Academia Sinica, Taipei 106, Taiwan}



\date{\today}

\begin{abstract}
We consider the thermodynamic potential $\Omega$ of an $N$ component
Fermi gas with a short range interaction
obeying SU($N$) symmetry.
We analyze especially the non-analytic part of $\Omega$ in
the temperature $T$ at low $T$.
We examine the temperature range where one can observe
this $T^4 \ln T$ contribution and discuss how it can be
extracted experimentally.

\end{abstract}


\maketitle



\section{Introduction}

For a non-interacting Fermi gas, textbooks \cite{LL1,AM} teach us that the
specific heat at low temperatures $T$ is a power series involving
only odd powers of $T$, as can easily be shown by the Sommerfeld
expansion.
Correspondingly, the grand thermodynamic potential $\Omega$
as a function of the temperature $T$, is thus
a power series in $T$ with only even powers.
 Interestingly, this result is qualitatively modified
for an interacting Fermi system, even with short range interactions
in two or three dimensions.
Though the first term in the expansion for the specific heat in $T$
indeed starts with
$T$, (for $\Omega$, a constant and a $T^2$ term),
the next term is now widely believed (see references below)
 to be of the form
$T^3 \ln T$
(correspondingly for $\Omega$, $T^4 \ln T$)
for three dimensions, thus not even analytic in $T$.
Thoroughly understanding these non-analytic terms
 is crucial in order to distinguish them from those arising in
 non-Fermi liquid phases or near
 quantum critical points \cite{NFL}.

Historically, the study of this $T^3 \ln T$ term was
motivated by the experimental observation in normal liquid $^3$He
that the specific heat cannot be fitted by the Sommerfeld
expression \cite{T3expr}.   On the theory side, this
non-analytic term can be understood to be due to
the presence of bosonic excitations (interacting particle-hole
pairs) in the system, even though these excitations
are not necessarily propagating but can be overdamped
(see, e.g. Refs. \cite{DE66,BS66}).
More interestingly, for the $T^3 \ln T$ term at least,
the result can be obtained via a proper
extension of the original
 Landau Fermi liquid theory \cite{Landau1,Landau2,LL,PN,Nozieres},
and the coefficient of this
$T^3 \ln T$ term can be entirely expressed in terms
of scattering amplitudes between Landau quasiparticles on the Fermi
surface \cite{Pethick73,BP}.
The non-analytic behavior in specific heat or thermodynamic potential
is a consequence of the non-analytic behavior of the density and spin
susceptibilities of the system at finite frequencies and
wavevectors \cite{Chubukov06}.

The theories of Refs. \cite{Pethick73,BP,Chubukov06} yield results for
this term that are in reasonable agreement with experiments \cite{T3expr}
in $^3$He.  However, precise statements are difficult to make due to
some uncertainties in the interacting parameters in this system
 \cite{Pethick73,BP,Chubukov06}.
This $T^3 \ln T$ term has also been studied in heavy fermion materials
such as UPt$_3$ and UAl$_2$ \cite{Coffey86,deVisser87}.
However, there
the interaction parameters are even less known than those in $^3$He.
Therefore it is highly desirable to have another system where these
theories can be tested.

In this paper, we analyze the thermodynamic potential of
 interacting SU($N$) Fermi gases
such as  $^{173}$Yb and $^{87}$Sr which are
available now in cold atom experiments
\cite{Yb,Sr,Yb-m,Sr-m1,Sr-m2,Pagano14,Zhang14}
(see also the review \cite{Cazalilla14}).
These $N_c$ components
(we use $N_c$ rather than $N$ here to avoid possible confusion with the number of
particles.)
represent the different choices
of hyperfine spin sublevels $m_f$ available to the atoms.
$m_f = -5/2, ... , 5/2$ for $^{173}$Yb and
$m_f = -9/2, ..., 9/2$ for $^{87}$Sr.
We would like to in particular examine whether these
systems can serve as candidates
to test these theories.   Both the number of components $N_c$,
($N_c$ can vary from $1$ to $6$ in $^{173}$Yb
and $1$ to $10$ in $^{87}$Sr)
and the density (hence the dimensionless coupling constant
defined below) can be varied in experiments.
(The former is possible due to the SU($N$) symmetry of
the interparticle interaction \cite{Pagano14,Zhang14}.)
For a sufficiently large cloud of the gas where the local density approximation
can be taken, the pressure $P$ of the gas
(which is equal to $-\Omega$ per unit volume)
can be deduced from the axial density \cite{trap-t1}.
Since the effective chemical potential varies across
the trap, analysis of this
data can then produce the grand thermodynamic potential
$\Omega$ as a function of the chemical potential $\mu$.
If the temperature can also be measured, then
the function $\Omega (\mu, T) $ can be obtained and compared with theory.
These types of studies have already been carried out extensively for many systems,
 including
two-component resonant Fermi gases
\cite{trap-ex1,trap-ex2,trap-ex3,trap-ex4,trap-ex5},
 one-component interacting Bose gas
 \cite{trap-ex-b1}, and
we expect that the same can be done for the $^{173}$Yb and $^{87}$Sr systems
eventually.
Previously we have investigated theoretically the Fermi liquid
properties of this SU($N$) Fermi gas at zero temperature
\cite{Yip14}, and we here extend our study to finite temperatures,
limiting ourselves to three dimensions in this paper.  While
the theories in Refs. \cite{Pethick73,BP,Chubukov06} pointed out
the existence of a $T^3 \ln T$ term in the specific heat
and thus a $T^4 \ln T$ term in $\Omega(\mu,T)$,
these calculations have not been verified numerically to
the best of our knowledge.  More importantly, they also
offer us no hint on the temperature range where one can
find such non-analytic behavior.  We here evaluate the contributions to
$\Omega(\mu,T)$ term by term numerically at arbitrary
temperatures which then allow us to answer this question.

In principle the non-analytic terms in the thermodynamic potential
can also be investigated for the resonant two-component system
or multi-component Fermi system without SU($N$) symmetry,
but we shall discuss how the variable $N_c$
may offer us some advantage.

\section{The Thermodynamic Potential $\Omega(\mu,T)$}

We present here the evaluation of the thermodynamic potential
$\Omega(\mu,T)$ of a $N_c$ component interacting Fermi gas
as a function of the chemical potential $\mu$ and temperature $T$.
$\mu$ is taken to be the same for all $N_c$ components.
The interparticle interaction is characterized by
a positive
s-wave scattering length $a$,
which is the same irrespective of the hyperfine spin sublevels $m_f$'s of the
fermions participating in the interaction
\cite{Pagano14,Zhang14}.
Note that the pressure $P$ is just $-\Omega/V$, where $V$ is
the volume.  We evaluate $\Omega(\mu,T)$ up to second order in $a$,
expressed as a power series in the dimensionless parameter $k_{\mu} a$,
where $k_{\mu} \equiv (2 M \mu)^{1/2}$ with $M$ the mass
of an atom.  $k_{\mu}$ would be equal to the Fermi momentum
in the special limit of zero temperature and in the absence of
interactions.

The zeroth order term of $\Omega(\mu,T)$ in $a$ is simply
that of the free gas, $\Omega_0(\mu,T)$, and hence given by
\begin{widetext}
\bdm
\Omega_0 (\mu,T) = N_c \sum_{\vec k} \left[ (\epsilon_k^0 - \mu) f_k^0
+ T  ( f_k^0 \ln f_k^0 + (1-f_k^0) \ln (1-f_k^0) ) \right]
\edm
\end{widetext}
where $\epsilon_k^0 \equiv \frac{k^2}{2M}$ is just
the kinetic energy,
\bdm
f_k^0 (\mu,T) \equiv \frac{1 } {  \exp ( \frac{ \epsilon_k^0 - \mu}{T}) + 1 }
\edm
the Fermi function, both of a free particle of wavevector $\vec k$.
$k$ is the magnitude of $\vec k$.
Here we have already used the fact that
$f^0_{\vec k, \alpha}, f^0_{\vec k, \beta}, ... $, the distribution
functions for species $\alpha, \beta, ...$ are all
given by $f^0_k (\mu,T)$, since
we have assumed equal chemical potentials for all species.
We have thus
\be
\Omega_0 (\mu,T) = V N_c  T \int \frac{d^3 k}{(2 \pi)^3}
\ln (1 - f_k^0)
\label{omega0}
\ee

\begin{figure}
\begin{center}
\includegraphics[width=69mm]{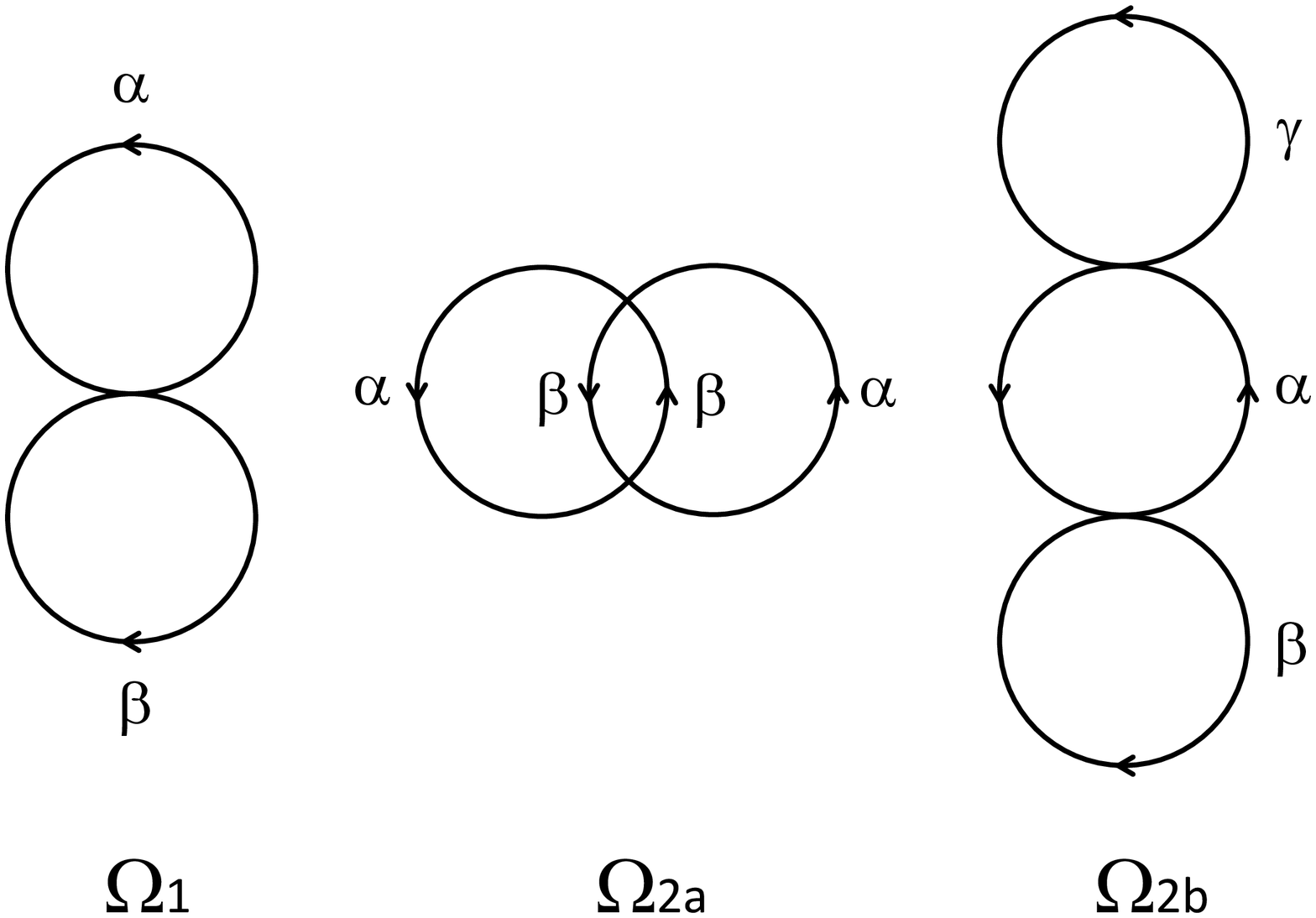}
\caption{
Feynman diagrams for the first $\Omega_1$ and second order contributions
$\Omega_{2a}$, $\Omega_{2b}$
in interaction to the thermodynamic potential.
}
\label{fig1}
\end{center}
\end{figure}

The Feynmann diagrams for the first and second order terms in $a$ are
shown in Fig 1.
The first order term $\Omega_1 (\mu,T)$ is simply
$ \frac{4 \pi a}{M V} \sum_{\alpha > \beta} N^0_\alpha N^0_{\beta}$,
where
$N^0_{\alpha}$, $N^0_{\beta}$ are the total number of
particles for species $\alpha$ and $\beta$ respectively.
In this expression, the sum over components is
restricted to different species since,
for short range interactions, the contributions
from direct and exchange interaction cancel for
identical species.  Furthermore, we have used the fact
that, at this order, it is sufficient to use
the particle numbers $N^0_{\alpha,\beta}$ at zeroth order.
 We thus have
\be
\Omega_1 (\mu,T) = V \frac{ N_c (N_c - 1)}{2}
\frac{ 4 \pi a}{M} n^0_{\alpha} n^0_{\beta}
\label{omega1}
\ee
where there is no sum over $\alpha,\beta$ in the above formula,
and $n^0_{\alpha}(\mu,T)$ is simply the number density of the $\alpha$
component, given by $\int \frac{d^3 k}{(2\pi)^3} f_k^0 $.

There are two diagrams to second order in $a$.  The first
one,  which we shall denote as $\Omega_{2a}$
and is depicted in the middle of Fig. \ref{fig1}, is given by
\begin{widetext}
\be
\Omega_{2a}(\mu,T)  = - \frac{ N_c (N_c - 1) }{2}
\left(\frac{ 4 \pi a}{M V}\right)^2
\sum_{\vec k_1, \vec k_2, \vec k'_1}
\frac{ f^0_{\vec k_1, \alpha} f^0_{\vec k_2, \beta}
( f^0_{\vec k_1', \alpha} + f^0_{\vec k_2', \beta} )}
{\frac{ k_1^2 + k_2^2 - k'_1{}^2 - k'_2{}^2}{2M}}
\label{Omega2a}
\ee
where $\alpha,\beta$ again are not summed,
and $\vec k'_2 \equiv \vec k_1 + \vec k_2 - \vec k'_1$.
$\Omega_{2a}$ is the only term in the thermodynamic potential, up
to this order in $a$,
which is responsible for modifications of the
physical properties of the system that
cannot be regarded as just a chemical potential shift
due to interaction.
This diagram is also responsible for the induced interaction
among Landau quasiparticles studied in, e.g., Refs. \cite{Yu12,Yip14}.
The second diagram, which we shall denote as $\Omega_{2b}$ and is
depicted on the right part of Fig. \ref{fig1},
can be considered
as a Hartree correction to the diagram for $\Omega_1$:  for example,
one can regard the line labeled by $\gamma$
is simply giving a constant energy shift
$\delta \epsilon = \frac{ 4 \pi a}{M} n^0_{\gamma}$
to the propagator labeled by $\alpha$.
Noting the combinatorial factor of $1/ 2!$ for second order interaction
diagrams, the part that is of order $a^2$ is thus
$\frac{1}{2} \frac{4 \pi a}{M V}
 \sum_{\alpha > \beta} (N_{\alpha} - N^0_{\alpha}) N^0_{\beta}$
where the difference
$(N_{\alpha} - N^0_{\alpha})/V$ is given by
\bdm
\int \frac{ d^3 k}{(2 \pi)^3}
\left( \frac{1 } {  \exp ( \frac{ (\epsilon_k^0+\delta \epsilon- \mu}{T}) + 1 }
-
\frac{1 } {  \exp ( \frac{ \epsilon_k^0 - \mu}{T}) + 1 } \right)
= - \frac{ \partial n^0_{\alpha}}{\partial \mu} (\delta \epsilon)
\qquad .
\edm
Summing over possible choices of $\gamma$ finally gives us
\be
\Omega_{2b} (\mu,T)
= V \frac{ N_c (N_c - 1)^2}{2}
\left( \frac{ 4 \pi a}{M} \right)^2
\left( - \frac{ \partial n^0_{\alpha}}{\partial \mu} \right)
n^0_{\beta} n^0_{\gamma}
\label{Omega2b}
\ee
where again $\alpha$, $\beta$, $\gamma$ are not summed.

We therefore have,
up to second order in $a$,
\be
\Omega(\mu,T) = V N_c \frac{k_\mu^3}{6 \pi^2}
\frac{k_{\mu}^2}{2M}
\left\{ \tilde \omega_0 + ( k_{\mu} a) \tilde \omega_1
 + ( k_{\mu} a)^2 (\tilde \omega_{2a} + \tilde \omega_{2b})
 \right\} \ ,
\label{Omega}
\ee
\end{widetext}
with
\bea
\tilde \omega_0 &=& \ \omega_0 \nonumber \\
\tilde \omega_1 &=& (N_c -1 ) \ \omega_1 \nonumber \\
\tilde \omega_{2a} &=& (N_c -1 ) \  \omega_{2a}  \label{Ndep} \\
\tilde \omega_{2b}  &=& (N_c - 1)^2 \ \omega_{2b} \nonumber
\eea
where $\omega_0$, ..., $\omega_{2b}$ are $N_c$ independent dimensionless
functions of $\mu$, $T$ and hence only of $t \equiv T/\mu$,
and $\omega_0$ originated from $\Omega_0$,
$\omega_1$ from $\Omega_1$, etc.
We shall provide the explicit expressions for $\omega_0$, ..., $\omega_{2b}$
later after we discuss the zero temperature limit.

At $T=0$, we easily find, using $n_{\alpha}(\mu,0) = k_{\mu}^3/6 \pi^2$,
$\partial n_{\alpha}(\mu,0)/\partial \mu = M k_{\mu}/2 \pi^2$,
\be
\omega_0(0) = - \frac{2}{5} \ ,
\label{omega0T0}
\ee
and
\be
\omega_1(0) =  \frac{2}{3 \pi} \ .
\label{omega1T0}
\ee
$\omega_{2a}$ is given by a rather complicated integral but
has already been evaluated before in the literature, as
the same integral appears in the energy of a
two-component Fermi gas to second order in $a$,
see for example \S 6 of Ref. \cite{LL}.
We find then
\be
\omega_{2a}(0) = \frac{4}{35} \frac{ ( 11 - 2 \ln 2)}{\pi^2}
\approx 0.11132 \ .
\label{omega2aT0} \\
\ee
$\omega_{2b}$ can be easily found to be
\be
\omega_{2b}(0) = -\frac{4}{3 \pi^2}
\approx -0.1351 \ .
\label{omega2bT0}
\ee

The total number of particles $N_{\rm tot}$ can be found via
$- \partial \Omega / \partial \mu$, and
defining the Fermi momentum $k_F$ via
$N_{\rm tot} = V N_c \frac{k_F^3}{6 \pi^2}$
(with $N_{\rm tot}$ the total number of particles at $T=0$) gives us
\bea
k_F &=& k_{\mu} \left\{
1 - 3 (k_{\mu} a) \tilde \omega_1 -
\frac{7}{2} (k_{\mu} a)^2 \tilde \omega_2  \right\}^{1/3} \label{kF} \\
&\approx& k_{\mu} \left\{
  1 -  (k_{\mu} a) \tilde \omega_1 -
  (k_{\mu} a)^2 [ \frac{7}{6} \tilde \omega_2 + \tilde \omega_1^2] \right\}
  \nonumber
\eea
where $\tilde \omega_2 = \tilde \omega_{2a} + \tilde \omega_{2b}$.
Using the relation $E = \Omega + \mu N_{\rm tot}$ for the total energy $E$
at zero temperature and eliminating $\mu$ in favor of $k_F$,
one can check that (see Appendix \ref{AppA}) our expressions above
reproduce the result for $E$ given in the literature
(e.g. Refs. \cite{LL,FW1,FW2}).

At finite temperatures, the dimensionless functions
$\omega_{0}$, ... , $\omega_{2b}$ are given by
\be
\omega_0 (t) = 3 \ \frac{T}{\mu}\ \frac{1}{k_{\mu}^3}
\int_0^{\infty} d k k^2
\ln ( 1 - f_k^0) \ ,
\label{o0gt}
\ee
\be
\omega_1 (t)= \frac{2}{3 \pi}
\left( \frac{n_{\alpha}^0(\mu,T)}{n_{\alpha}^0(\mu,0)} \right)^2 \ ,
\label{o1gt}
\ee
\be
\omega_{2a} (t) = - \frac{3 \times 2^6 \times \pi^4}{k_{\mu}^7}
\int_{\vec {k}_1} \int_{\vec k_2} \int_{\vec k'_1}
\frac{ f_{\vec k_1} f_{\vec k_2} ( f_{\vec k'_1} + f_{\vec k'_2} )}
{k_1^2 + k_2^2 - k'_1{}^2 - k'_2{}^2  }
\label{o2agt}
\ee
(where we have introduced the short-hand
$\int_{\vec k} \equiv \int \frac{d^3 k}{(2 \pi)^3}$),
and
\be
\omega_{2b} (t) = -\frac{4}{3 \pi^2}
\left( \frac{n_{\alpha}^0(\mu,T)}{n_{\alpha}^0(\mu,0)} \right)^2
\left( \frac{ \partial n^0_{\alpha} (\mu,T)/\partial \mu}
{\partial n^0_{\alpha} (\mu,0)/\partial \mu} \right)  .
\label{o2bgt}
\ee

Below we discuss the low temperature expansion of $\Omega (\mu,T)$.
It is convenient first to ignore the contribution from $\omega_{2a}$,
that is, we include the Hartree-Fock diagrams only.
We shall call  this result $\Omega^{\rm HF} (\mu,T)$.
The low temperature formulas for $\omega_{0,1,2b}$ can
be easily obtained by standard Sommerfeld expansion.
We get
\be
\omega_0(t) = - \frac{2}{5} \left[ 1 +
\frac{5 \pi^2}{8} t^2 - \frac{7 \pi^4}{ 384} t^4 +
...\right]  \ ,
\label{o0t}
\ee
and with the help of
\bdm
\frac{n^0_{\alpha}(\mu,T)}{n^0_{\alpha}(\mu,0)} =   1 + \frac{\pi^2}{8} t^2 +
\frac{7 \pi^4}{640} t^4 + ...  \ ,
\edm
\bdm
\frac{\partial n^0_{\alpha}(\mu,T)/\partial \mu}
{\partial n^0_{\alpha}(\mu,0)/\partial \mu}  =
 1 - \frac{\pi^2}{24} t^2 -
 \frac{7 \pi^4}{ 384} t^4 + ...  \ ,
\edm
also
\be
\omega_1(t) =  \frac{2}{3 \pi} \left[ 1 +
\frac{\pi^2}{4} t^2  +
\frac{3 \pi^4}{80} t^4  +... \right] \label{o1t}
\ee
and
\be
\omega_{2b}(t) = - \frac{4}{3 \pi^2} \left[ 1 +
\frac{5\pi^2}{24} t^2 + \frac{17 \pi^4}{1920} t^4
+ ... \right] \label{o2bt}
\ee
where $t \equiv T/\mu$ and the higher order terms not displayed
here are $t^6$, $t^8$ etc.

The above implies, to order $T^2$,
\begin{widetext}
\be
\Omega^{\rm HF} (\mu,T) =
\Omega^{\rm HF} (\mu,0)
- V N_c \frac{M k_{\mu}}{2 \pi^2} \frac{\pi^2 T^2}{6}
\left[ 1 - (N_c - 1) \frac{ 2 k_{\mu} a}{3 \pi}
+ (N_c - 1)^2 \frac{10}{9} \left(\frac{k_{\mu} a}{\pi} \right)^2 \right] \ .
\label{HF}
\ee
\end{widetext}

This result is in accordance with the expectation
from Fermi liquid theory \cite{LL,PN}, though
with interactions now restricted to Hartree-Fock.
In this theory, the specific heat should be linear in
$T$ at low temperatures, and is given by
$V N_c \frac{M k_F^{\rm HF}}{2 \pi^2} \frac{\pi^2 T}{3}$.
Note that the density of states for each fermion component
 $\frac{M k_F^{\rm HF}}{2 \pi^2}$  that enters here is related to the Fermi
wavevector $k_F^{\rm HF}$ for the corresponding particle
density {\em at zero temperature}.
Since we are using the chemical potential as an independent
variable, $k_F^{\rm HF}$ is given by the corresponding
Hartree-Fock value, thus by
Eq. (\ref{kF}) but with the contribution from $\tilde \omega_{2a}$ dropped.
Indeed,
using Eqs. (\ref{omega0T0}),(\ref{omega1T0}) and (\ref{omega2bT0}),
we obtain
\be
k_F^{\rm HF} = k_{\mu} \left[
 1 - (N_c - 1) \frac{ 2 k_{\mu} a}{3 \pi}
+ (N_c - 1)^2 \frac{10}{9} \left(\frac{k_{\mu} a}{\pi} \right)^2 \right] \ .
\ee
Together with $S^{\rm HF} = - \partial \Omega^{\rm HF}/\partial T$,
and noting that to linear order in $T$,
the entropy $S^{\rm HF} (\mu,T)$ of the system is given by the same expression
as the specific heat, we
verify our claim above.

$\omega_{2a}$, in contrast to the other terms discussed above,
is {\em not} expected to be analytic in $t$.
Rather, we anticipate
\be
\omega_{2a}(t) = \frac{4}{35\pi^2} (11 - 2 \ln 2)
+B_{22} t^2 + B_{23} t^4 \ln t +B_{24} t^4 + ...
\label{Bdef}
\ee
The first term was already given in Eq. (\ref{omega2aT0}).

Let us first discuss $B_{22}$.  With similar discussions
for the Hartree-Fock contributions above,
we expect from Fermi liquid theory that the thermodynamic
potential, up to $T^2$, is given by
\be
\Omega (\mu,T) = \Omega (\mu,0) - V N_c
\frac{M^* k_F}{2 \pi^2} \frac{\pi^2 T^2}{6}
\label{FL}
\ee
with now $k_F$ given by Eq. (\ref{kF}), and
$M^*$ the effective mass of the quasiparticles.
$M^*$ is available from standard text with rather straight-forward
extension \cite{Yip14} to our SU($N$) system.  We have
\be
M^*/M = 1 + (N_c - 1) \frac{8}{ 15 \pi^2} ( 7 \ln 2 - 1)
(k_{\mu} a)^2
\label{M*}
\ee
where we have already taken the liberty that, at this order,
we can simply use $k_{\mu}$ instead of $k_F$ in the last term.
Eqs. (\ref{FL}) and (\ref{HF}) imply that we expect
\be
\Omega_{2a}(\mu,T) = \Omega_{2a}(\mu,0)
- V N_c \frac{M^* k_F - M k_F^{\rm HF}}{2 \pi^2} \frac{\pi^2 T^2}{6} \ .
\label{O2a}
\ee
Using Eqs. (\ref{M*}) and (\ref{kF}),
we obtain, to second order in $a$,
\bdm
M^* k_F - M k_F^{\rm HF} =
 M k_{\mu} (N_c - 1) \left( \frac{k_{\mu} a}{\pi} \right)^2
\ 2 [ 2 \ln 2  - 1 ] \ .
\edm
Eq. (\ref{Omega}) together with Eq. (\ref{Bdef}) show that we
anticipate
\be
B_{22} = - \ln 2 + 0.5 \approx - 0.1936 \ ,
\label{B22}
\ee
a value which we shall verify independently below.

The term proportional to $B_{23}$ is the first non-analytic
contribution to $\Omega_{2a}(\mu,T)$ and hence $\Omega(\mu,T)$
at low temperatures.  Theories presented in Refs.
\cite{Pethick73,Chubukov06} provided formulas
for this quantity, and
their results are in agreement with each other.
In  Ref. \cite{Pethick73}, Eq. (22), the non-analytic contribution to
the entropy $S$ was written as, for the two-component system
$N_c = 2$,
\be
\Delta S = - V \frac{\pi^4}{20} n_{\rm tot} B^s
\left( \frac{T}{T_F} \right)^3 \ln \left( \frac{T}{T_F} \right)
\label{DS}
\ee
where $n_{\rm tot}$ is the total density, $T_F$ the Fermi temperature.
To our required accuracies we can put $n_{\rm tot} = 2 k_{\mu}^3/ 6 \pi^2$
($N_c = 2$), and
replace $T_F$ by $\mu$. $B^s$ is a quantity that can be expressed
in terms of scattering amplitudes between particles.
To second order in these amplitudes, we have, via Eq. (65) in Ref. \cite{Pethick73},
\be
B^s = -\frac{1}{2} \left[ (A_0^s)^2 + 3 (A_0^a)^2 \right]
\ee
where
$A_0^{s}, A_0^{a}$ are the angular-averaged scattering amplitudes
symmetric and antisymmetric respectively with respect to spins.
To lowest order in $a$, they are in turn given by
$A_0^s = - A_0^a = \frac{2 k_{\mu} a}{\pi}$.
The same $\Delta S$ can be obtained from Ref. \cite{Chubukov06}
by combining their Eqs. (39), (11) and (12) with their $U$
replaced by $\frac{4 \pi a}{M}$.
On the other hand,
$\Delta S$ can be obtained from $- \partial \Omega/ \partial T$,
noting that it originates the $B_{23}$ term of Eq. (\ref{Bdef})
of $\omega_{2a}$ only.
We get
\be
\Delta S = - V N_c (N_c - 1) \frac{k_{\mu}^3}{ 6 \pi^2}
(k_{\mu} a)^2 ( 4 B_{23}) t^3 \ln t \ .
\label{DS2}
\ee
Comparison between Eqs. (\ref{DS}) and (\ref{DS2}) gives
\be
B_{23} =  - \frac{\pi^2}{10} \approx -0.9869  \ ,
\label{B23}
\ee
a value which
we shall check also later.

\begin{figure}
\begin{center}
\includegraphics[width=69mm]{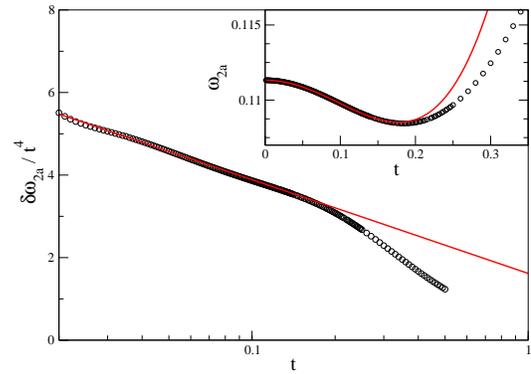}
\caption{
(Color online)
Inset: $\omega_{2a}(t)$ as a function of reduced temperature $t \equiv T/\mu$.
Main figure:
$\delta \omega_{2a}(t)$ divided by $t^4$, plotted as a function of
$\ln t$.  Here
$\delta \omega_{2a}(t) \equiv
\omega_{2a} - B_{20} - B_{22} t^2$, that is,
$\omega_{2a}(t)$ after subtraction of the lower order
analytic terms in $t$.
The values of $B_{20}$ and $B_{22}$ used in
this subtraction are $0.1113$ and $-0.1936$.
The straight line corresponds to $B_{23}=-0.9869$.  The
fit also gives $B_{24}$ defined in Eq. (\ref{Bdef}) to be
approximately $1.62$.
\\
}
\label{fig2}
\end{center}
\end{figure}

Now we present our numerical results.
We first consider $\omega_{2a} (t)$, presented in the inset of Fig. \ref{fig2}.
Our numerical results for this quantity at low temperatures agree
with what we expect from Eq. (\ref{Bdef})
with $B_{22}$ given in Eq. (\ref{B22}).   Subtracting these lower order
 (constant and $t^2$) analytic terms and defining the resultant quantity to be
$\delta \omega_{2a} (t)$, the plot of $\delta \omega_{2a}/t^4$
as a function of  $\ln t$ is given in the main part of Fig. \ref{fig2}.
The lower temperature data show clearly a $t^4 \ln t$
contribution to $\omega_{2a}(t)$, applicable for $t$ up
to $\approx 0.2$, where then we find deviation from
Eq. (\ref{Bdef}) due to contributions from higher order
terms in $t$ (which likely also contain further non-analytic contributions).
The slope of this plot gives $B_{23}$ also in good agreement
 Eq. (\ref{B23}).  There are some
deviations from the straight-line for very low temperatures
due to numerical inaccuracies from the subtraction.
The fit also gives us $B_{24} \approx 1.62$.

\begin{figure}
\begin{center}
\includegraphics[width=69mm]{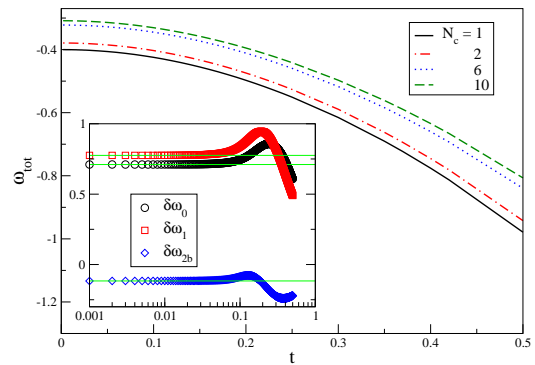}
\caption{
(Color online)
$\omega_{tot}$
\{$\equiv \Omega(\mu,T) / [V N_c k_{\mu}^5/ (12 \pi^2 M)]$\}
as a function of reduced temperature $t \equiv T/\mu$, with $k_{\mu} a = 0.1$.
Inset: $\delta \omega_{0,1,2b}(t) /t^4$ as functions of
$\ln t$. $\delta \omega_{0,1,2b}(t)$ are defined as
$\omega_{0,1,2b}(t)$ after subtraction of their
lower order (constants and $t^2$)
analytic terms  (c.f. Fig \ref{fig2}).
}
\label{fig3}
\end{center}
\end{figure}

Fig. \ref{fig3} shows an example for
the total thermodynamic potential $\Omega(\mu,T)$, in units
of $V N_c k_{\mu}^5/ (12 \pi^2 M)$ (see Eq. (\ref{Omega})) for
various values of $N_c$.  The non-analytic contributions
are not directly evident from this plot.
 The inset shows the
behavior of the analytic contributions $\omega_{0,1,2b}(t)$
plotted in analogous manner to the main Fig. \ref{fig2},
that is, after subtraction of the lower order $t$ terms
and divided by $t^4$.
Since these quantities are power series in $t^2$, after
these subtractions they become roughly constants at low
temperatures in this plot.  Their intersections with the $y$ axis
give values that are in full agreement with
the $t^4$ coefficients in Eqs. (\ref{o0t}-\ref{o2bt}).
Deviations from the horizontal lines are due to
contributions from higher order ($t^6$, $t^8$, ...) terms.
We see that they become significant for $t \gtrsim 0.1$.

\begin{figure}
\begin{center}
\includegraphics[width=69mm]{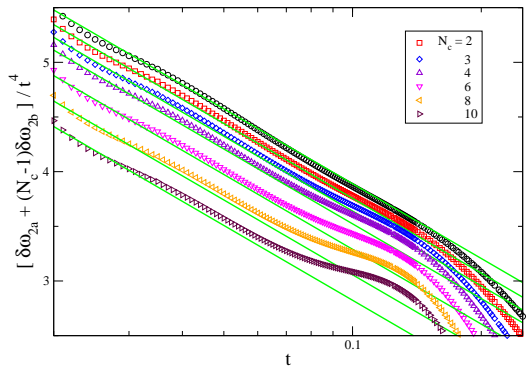}
\caption{
(Color online)
$[\delta \tilde \omega_{2a}(t) + \delta \tilde \omega_{2b}(t)]/(N_c - 1)t^4
=[\delta \omega_{2a}(t) + (N_c -1 )\delta \omega_{2b}(t)]/t^4    $
as a function of  $\ln t$.
$\delta \tilde \omega_{2a}(t) + \delta \tilde \omega_{2b}(t)$
is defined as $\tilde \omega_{2a}(t) + \tilde \omega_{2b}(t)$
after subtraction of their lower order analytic terms.
Circles $\circ$ represent the  limit $N_c \to 1$
hence $\delta \omega_{2a}(t) /t^4 $.
}
\label{fig4}
\end{center}
\end{figure}

If experimentally the pressure and hence $\Omega(\mu,T)$ can
be measured for various $\mu$, $T$ and $N_c$, one can
normalize this quantity to
$V N_c k_{\mu}^5/ (12 \pi^2 M)$, extract the coefficients of
$k_{\mu} a$ and $(k_{\mu} a)^2$ and obtain the quantities
$\tilde \omega_{0,1}$ and $\tilde \omega_{2a} + \tilde \omega_{2b}$
defined in Eq. (\ref{Omega}).
One can then fit $\tilde \omega_{2a}(t) + \tilde \omega_{2b}(t)$
at low temperatures to obtain the $t=0$ value and a $t^2$ contribution.
Subtracting these lower order analytic terms and let us define the
resulting quantity to
be $\delta \tilde \omega_{2a}(t) + \delta \tilde \omega_{2b}(t)$
[which
should then be given by $(N_c -1 ) \delta \omega_{2a} + (N_c -1)^2 \delta \omega_{2b}$].
This quantity, after division by $(N_c -1) t^4$, would
behave as what is plotted in Fig. \ref{fig4} for various $N_c$'s.
For a given $N_c\ge 2$ (note that $N_c = 1$ gives only
a non-interacting gas) there would be a $\ln t$ contribution.
The range where this $\ln t$ would be evident actually decreases
with $N_c$, and even for $N_c = 2$ is restricted to $t < 0.1$,
as compared with $\sim 0.2$ for $\omega_{2a}(t)$ in Fig. \ref{fig2}.
This is due to the contribution from the ``bump" near $t \sim 0.1$
arising from the $t^6, t^8, ...$ contributions we described
for $\delta \omega_{2b}(t)$ for the inset of Fig. \ref{fig3}.
However, if data for
various $N_c$'s are available, one can in principle
extrapolate the data at a given $t$ to $N_c = 1$ and
obtain the non-analytic term $\delta \omega_{2a}$.
Note that Fig. \ref{fig4} implies that, at large $N_c$,
the non-analytic contribution from $\tilde \omega_{2a}$ becomes
less and less important as compared with $\tilde \omega_{2b}$,
in accordance with the expectation that at large $N_c$,
the thermodynamic potential is more and more mean-field like
(see Eq. (\ref{Ndep})).

\section{Discussions}\label{Sec:Con}

In principle the non-analytic contribution to $\Omega$ can
also be studied for a two-component resonant Fermi gas
\cite{trap-ex1,trap-ex2,trap-ex3,trap-ex4,trap-ex5},
at temperatures above the superfluid transition temperature $T_c$.  At small
and negative scattering length $a$, the transition temperature
is small and there would still be a temperature range
where the $T^4 \ln T$ term should be observable.  The
advantage of studying this system is that $k_{\mu} a$ can be
varied over a wide range, and we can
study the higher order contributions in $k_{\mu} a$ not
analyzed in the present paper, though one has to
stay sufficiently above $T_c$ so that pairing fluctuations
would not introduce complications.
One can also use the ``upper" $a>0$ branch of the
Feshbach resonance at magnetic fields where the stability
of the gas is not an issue.   An advantage of this
case is that the higher order interaction terms,
not studied in this paper, may give rise to an enhancement
for the non-analytic term similar to what occurs in
$^3$He \cite{Pethick73}.
A disadvantage however is
that we only have $N_c = 2$ and the extrapolation procedure
described near the end of the last section is not available.
The $T^4 \ln T$ term would also be present for an
interacting Fermi gas without SU($N$) symmetry, with
again no extrapolating procedure in the component number feasible.

Experimentally, the density $n (\mu,T)$ can also be measured. Since
$n (\mu,T) =  - \partial \Omega (\mu,T)/ \partial \mu$,
it is also non-analytic in $T$ with
a $T^4 \ln T$ contribution when $N_c \ge 2$.   The necessary formulas
can be straight-forwardly derived from the ones we gave
here.  They are listed in Appendix \ref{sec:density}
and the non-analytic terms can be extracted by a
similar analysis as we discussed in text for $\Omega (\mu,T)$.

The extraction of the non-analytic terms in the thermodynamic potential
or density seems demanding as very accurate experimental data
would be required.  On the other hand, these studies would
shed valuable new light on an old and interesting problem.

\section{Acknowledgement}

This work is supported by the Ministry of Science and Technology,
Taiwan, under grant number MOST-104-2112-M-001-006-MY3.

\appendix
\section{Zero Temperature }\label{AppA}

\newcounter{seq}

\newenvironment{seq}{\refstepcounter{seq}\equation}{\tag{B\theseq}\endequation}

Here we verify that Eq. (\ref{Omega}), together with
Eqs. (\ref{omega0T0})-(\ref{omega2bT0}), does yield the correct
 result for the energy $E$ given in the literature \cite{LL,FW1,FW2}.
The total number of particles $N_{\rm tot}$ can be obtained
by $- \partial \Omega / \partial \mu$, and so
\be
N_{\rm tot} = V N_c
\frac{k_{\mu}^3}{6 \pi^2}  \left\{
1 - 3 (k_{\mu} a) \tilde \omega_1 -
\frac{7}{2} (k_{\mu} a)^2 \tilde \omega_2  \right\}
\ee
hence, with $E = \Omega + \mu N_{\rm tot}$,
\be
E = V N_c
\frac{k_{\mu}^3}{6 \pi^2} \frac{k_{\mu}^2}{2M}  \left\{
\frac{3}{5} - 2 (k_{\mu} a) \tilde \omega_1 -
\frac{5}{2} (k_{\mu} a)^2 \tilde \omega_2  \right\}
\label{E}
\ee
$k_F$ was already obtained in Eq. (\ref{kF}).  Inverting
that equation, we obtain
\be
k_{\mu} = k_F \left[ 1 +  \tilde \omega_1 (k_F a)
  + ( \frac{7}{6} \tilde \omega_2 + 3 \tilde \omega_1^2) (k_F a)^2 \right]
\label{kmu}
\ee
At this stage, we can already verify the dependence of the
chemical potential $\mu$ on $k_F$, since $\mu \equiv k_{\mu}^2/2M$.
With the help of the zero temperature values of
$\omega_{1,2a,2b}$ in Eqs. (\ref{omega1T0}),(\ref{omega2aT0}) and
(\ref{omega2bT0}),
we get
\begin{widetext}
\be
\mu = \frac{k_F^2}{2M}
\left[ 1 +  (N_c -1 ) \frac{4}{3 \pi} (k_F a)
  + (N_c - 1) \frac{4}{15} \frac{ 11 - 2\ln 2}{\pi^2} (k_F a)^2 \right]
\ee
in agreement with, e.g., Ref. \cite{FW1}.
Note that the last term is proportional to $(N_c -1)$ and contributions
that are $(N_c -1 )^2$ mutually cancel.
Substituting Eq. (\ref{kmu}) into Eq.
(\ref{E}) and directly using the relation between $N_{\rm tot}$ and $k_F$ gives
us
\be
\frac{E}{N_{\rm tot}} = \frac{k_F^2}{2M}
\left[ \frac{3}{5} +  (N_c -1 ) \frac{2}{3 \pi} (k_F a)
  + (N_c - 1) \frac{4}{35} \frac{ 11 - 2\ln 2}{\pi^2} (k_F a)^2 \right]
\ee
\end{widetext}
in agreement with Refs. \cite{LL,FW1,FW2}.

\section{particle density $n(\mu,T)$ }\label{sec:density}

We give here the low temperature expansion for the density
$n(\mu,T)$.  We write it in a form similar to $\Omega(\mu,T)$ in text.
We have

\begin{widetext}
\be
n(\mu,T) = V N_c \frac{k_\mu^3}{6 \pi^2}
\left\{ \tilde \nu_0 + ( k_{\mu} a) \tilde \nu_1
 + ( k_{\mu} a)^2 (\tilde \nu_{2a} + \tilde \nu_{2b})
 \right\}
\label{n}
\ee
\end{widetext}
where
\bea
\tilde \nu_0 &=& \ \nu_0 \nonumber \\
\tilde \nu_1 &=& (N_c -1 ) \ \nu_1 \nonumber \\
\tilde \nu_{2a} &=& (N_c -1 ) \  \nu_{2a}  \label{nNdep} \\
\tilde \nu_{2b}  &=& (N_c - 1)^2 \ \nu_{2b} \nonumber
\eea
with
\be
\nu_0(t) =   1 + \frac{\pi^2}{8} t^2 +
\frac{7 \pi^4}{640} t^4 + ...  \ ,
\ee

\bea
\nu_1(t) &=&  \frac{2}{3 \pi} \left[ -3 -
\frac{\pi^2}{4} t^2  +
\frac{3 \pi^4}{80} t^4  +... \right]  \\
\nu_{2b}(t) &=&  \frac{4}{3 \pi^2} \left[ \frac{7}{2} +
\frac{5\pi^2}{16} t^2 - \frac{17 \pi^4}{3840} t^4
+ ... \right] \ ,
\eea
and
\begin{widetext}
\be
\nu_{2a}(t) =  - \frac{7}{2} \omega_{2a}(0)
 - \frac{3}{2} B_{22} t^2 + \frac{1}{2} B_{23} t^4 \ln t +
 (\frac{B_{24}}{2} + B_{23}) t^4 + ... \ ,
\ee
\end{widetext}
where $\omega_{2a}(0)$, $B_{22}$, $B_{23}$, $B_{24}$ are
the same coefficients that appeared in text for $\Omega(\mu,T)$.


\end{document}